\documentclass[aip,cha,numerical]{revtex4-1}
\usepackage{amsmath} %twocolumn,
\usepackage{graphicx}
\usepackage{amsfonts}% Include figure files
\usepackage{dcolumn}% Align table columns on decimal point
\usepackage{bm}% bold math
\usepackage{CJK}

\begin{document}
\begin{CJK*}{GB}{}
\title{Delay induced bifurcation of dominant transition pathways}

\author{Huijun Jiang \CJKfamily{gbsn} (½­»Û¾ü)}
\author{Zhonghuai Hou  \CJKfamily{gbsn}(ºîÖл³)}
\email{hzhlj@ustc.edu.cn}

\affiliation{ Department of Chemical Physics \& Hefei National Laboratory for Physical Sciences at the Microscale, University of
 Science and Technology of China,Hefei, Anhui 230026, China}

\begin{abstract}
We investigate delay effects on dominant transition pathways (DTP) between metastable states of stochastic systems. A modified version of the Maier-Stein model with linear delayed feedback is considered as an example. By a stability analysis of the {\lq on-axis\rq} DTP in trajectory space, we find that a bifurcation of DTPs will be induced when time delay $\tau$ is large enough. This finding is soon verified by numerically derived DTPs which are calculated by employing a recently developed minimum action method extended to delayed stochastic systems. Further simulation shows that, the delay-induced bifurcation of DTPs also results in a nontrivial dependence of the transition rate constant on the delay time. Finally, the bifurcation diagram is given on the $\tau-\beta$ plane, where $\beta$ measures the non-conservation of the original Maier-Stein model.
\end{abstract}
\begin{quotation}
The fluctuation-driven transition between metastable states is particularly relevant to many important events in physics, chemistry, biology, etc. Examples include chemical reactions, biological switches, nucleation processes, to list just a few. In many systems, the dynamics may involve delayed interactions due to limit transmission rate of matter, energy or information transport, or some kinds of feedback. Thus, it is of great interest to investigate the impact of delayed interactions on the transition behaviors of such systems. In this paper, we address this issues by investigation the effect of delayed interaction on the dominant transition pathways which is informative to explain the mechanism of fluctuation-driven transitions.
\end{quotation}

\maketitle
\section{Introduction}
Real dynamical systems are often subject to weak random perturbations, such as thermal noise at a nonzero temperature. It has been a common sense that small fluctuations can produce a profound effect on the long time dynamics by inducing rare but important events. For instance, fluctuations may result in transitions between metastable sets of deterministic dynamical system, which can be related to a large number of interesting phenomena in physics, chemistry and biology such as nucleation processes, chemical reactions, and biological switches.

In recent years, fluctuation-driven transitions (FDT) have gained great research attentions\cite{PRL93001783, PRE92003691, PhR53001505, PRL08140601, JCP08061103,  Ren2002, JSP04001577, CMS10000341, NON10000475}. One of the fundamental purposes of studying FDT is to explain how the transition occurs. Freidlin-Wentzell theory of large deviations provides one of the right frameworks to understand FDT\cite{Fre1998, Shw1995, Var1984}. When the amplitude of fluctuation is small, the distribution of trajectories which make transitions between metastable sets is often sharply peaked around a certain deterministic path or a set of paths. It then becomes very important to identify such dominant transition pathways (DTP) which can be highly informative to help elucidate the underlying mechanism of the FDT. Usually, the DTP tells how the transition happens step by step, identifies the transition state(s), and can also be used to derive other important quantities such as the transition rate of the FDT. In conservative systems, where there exists an underlying energy landscape \cite{JCP08061103, Ren2002, PhR53001505}, the DTP is actually the minimum energy path and is everywhere tangent to the potential force. In this case, the DTP must first approach a transition state which is usually a saddle point on the basin boundary of one attractor, and then runs along the unstable manifold of that point and enters the basin of attraction of the second attractor. The DTP before reaching the transition state is actually the time reversed heteroclinic orbit of the unperturbed system joining the attractor and the saddle point. In non-conservative systems in which detailed balance is absent, some new interesting phenomena have been found, e.g., symmetry breaking bifurcation of the optimal escape path can be observed\cite{PRL93001783}, an unstable fixed point \cite{PRE92003691} or an unstable limit cycle \cite{JSP04001577} can act as the transition state. Moreover, a complex transition paradigm containing a saddle point and two limit cycles as transition states was reported in the Lorenz system \cite{CMS10000341}. The DTP has also been used to explore the configuration space of systems with complicated structure \cite{NON10000475}, and study the nucleation process in the presence of shear in a two-dimensional Ginzburg-Landau equation \cite{PRL08140601}.

Nevertheless, most studies of DTP so far are limited at least in one sense, i.e., the state of system at one time can only influence and be influenced by its state at that same time. In fact, a variety of sources, such as the limit transmission rate of matter, energy or information transport, or some kinds of feedback, might allow events at one time to affect the state of the system at some later time. In these cases, time delayed variables and equations should be used to describe the dynamics. As we already know, delay models have been widely used to describe chemical kinetics \cite{JPC96008323}, neuronal networks \cite{Hak2002}, circadian oscillators \cite{EMB01000109, EMB97007146, EMB99004961}, physiological systems \cite{SCI87000287}, optical devices \cite{PRL80000709}, and so on. In addition, time delay can lead to a variety of interesting and important phenomena, such as delay-induced oscillation\cite{PNA05014593}, delay-induced excitability\cite{PRL05040601}, delay-induced oscillator death\cite{PRL98005109}. However, to the best of our knowledge, how delay would influence the FDT dynamics of a stochastic system, albeit its apparent importance, has not been studied yet.

In present paper, we have addressed such an issue by investigating the effect of time delay on the DTP in a modified version of the Maier-Stein model with linear delayed feedback. By an analysis using small delay approximation, we find that the DTP undergoes a bifurcation via transverse instability in the trajectory space when the delay time $\tau$ bypass a certain threshold value $\tau_c$. By extending a recently developed minimum action method \cite{PAM04000637} to this delayed stochastic system, we have also obtained  the DTPs by numerical simulations, which further confirms the analytical results. In addition, this bifurcation of DTP results in a nontrivial phenomenon of the FDT: The transition rate constant between two metastable states shows distinct dependence on the delay time below and above the bifurcation point. Finally, the bifurcation diagram is given on the $\tau-\beta$ plane, where $\beta$ stands for the non-conservation effect of the original Maier-Stein model.

\section{Analysis}
In general, a delayed stochastic system whose dynamics is determined by both the present state ${\bf{x}}(t)$ and the state ${\bf{x}}(t - \tau )$ with the delay time $\tau  > 0$ can be described as
\begin{align}
{\bf{\dot x}}(t) = {\bf{F}}({\bf{x}}(t),{\bf{x}}(t - \tau )) + \sqrt \varepsilon {\bf{\sigma}}({\bf{x}}){\bf{\eta }}(t),
\label{eq1}
\end{align}
\noindent where ${\bf{F}}({\bf{x}}(t),{\bf{x}}(t - \tau ))$ is a known drift vector field and ${\bf{\eta }}(t)$  is a set of independent Guassian white noises with zero mean and unit variance. $\varepsilon$ is a small positive number, and $\sigma({\bf{x}})$ is related to the diffusion tensor by $a = \sigma \sigma ^T$.

\begin{figure}
\centerline{\includegraphics*[width=0.7\columnwidth]{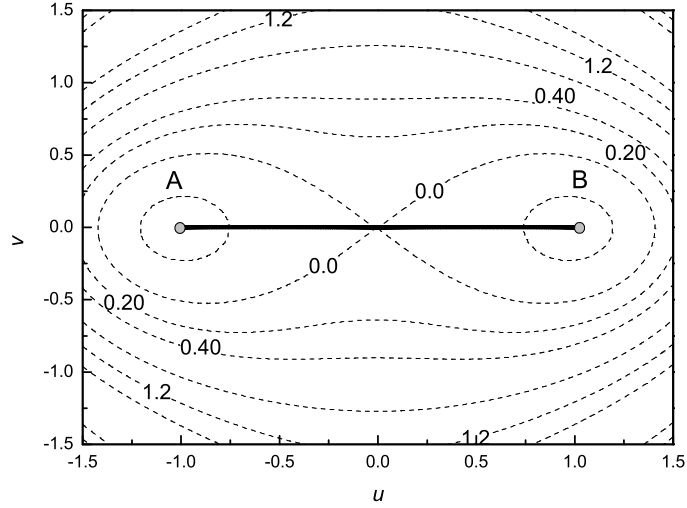}} \caption{ The potential field and the minimum energy path (the bold solid line) of the original Maier-Stein model.\label{Fig1}}
\end{figure}

To show the effect of delay, here we consider a modified version of the Maier-Stein model with state vector ${\bf{x}}=(u,v)$ as an example, whose linear term is modified to be a delayed feedback.
\begin{equation}\label{eq12}
\begin{split}
\dot u (t) &= u_{t-\tau}   - u^3  - \beta uv^2 + \sqrt \varepsilon {\bf{\eta }} ^u (t)   \\
\dot v (t) &=  - v_{t-\tau}   - u^2 v + \sqrt \varepsilon {\bf{\eta }} ^v (t)
\end{split}.
\end{equation}
\noindent When $\tau =0$, Eq.(\ref{eq12}) recovers to the original Maier-Stein model. In the absence of noise terms, it has two stable steady states at ${\bf{A}}=(-1,0)$, ${\bf{B}}=(1,0)$ and a saddle point at the origin $(0,0)$ for all values of $\beta>0$. In the presence of weak noise, however, both {\bf{A}} and {\bf{B}} become metastable states (MSS), and the FDT from one MSS to another is a rare event. $\beta$ reflects the non-conservation of the original model. For $\beta =1$, the drift field of the original system can be viewed as a gradient of a potential field, and the DTP from {\bf{A}} to {\bf{B}} (or vice versa) is actually the minimum energy path connecting {\bf{A}} and {\bf{B}} along the $u$ axis, which is shown in Fig.\ref{Fig1}.

In the delayed Maier-Stein model whose dynamical equation is governed by Eq.(\ref{eq12}), {\bf{A}} and {\bf{B}} are still the asymptotic fixed points of the system. By a simple linear stability analysis, we can find that {\bf{A}} and {\bf{B}} are stable for $\tau >0$. Similar to the case without delay, when small perturbation is present, FDT from one MSS to another is allowed. As this study focuses on the effect of delay, we will fix $\beta=1$ if not otherwise stated to avoid the influence of instantaneous non-conservative effects.

We now try to figure out the critical value of $\tau$ at which the on-axis DTP will be unstable. To this end, we expand Eq.(\ref{eq12}) in powers of $\tau$ using a small delay expansion around ${\bf{x}}(t)$ \cite{PRE99003970} as follows,
\begin{equation}\label{eq13}
\begin{split}
\dot u (t) &= (1-\tau) (u   - u^3  - uv^2) + \sqrt \varepsilon (1-\tau){\bf{\eta }} ^u (t)   \\
\dot v (t) &=  (1+\tau)(- v   - u^2 v) + \sqrt \varepsilon (1+\tau){\bf{\eta }} ^v (t)
\end{split},
\end{equation}
\noindent When weak noise presents, for a given transition path $\Phi=\{\Phi_0={\bf{A}},...,\Phi_T={\bf{B}}\}$ from MSS ${\bf{A}}$ to ${\bf{B}}$ over a finite time interval $T$, a Freidlin-Wentzell action functional $S_T [\Phi ]$ can be calculated by path intergral along $\Phi$. The main result of the Freidlin-Wentzell theory\cite{Fre1998} is that for sufficiently small $\varepsilon$, a probability can be assigned for path $\phi$ as $P(|\{{\bf{x}}\}-\Phi|<\epsilon)\approx \exp\{-S_T[\Phi]/\varepsilon\}$, where $\epsilon$ is a sufficiently small positive number. The probability suggests that DTP(s) is(are) the transition path(s) minimizing the action functional. For the on-axis DTP $\psi=\{\psi_0={\bf{A}},...,\psi_T={\bf{B}}, \psi^v=0\}$, the Freidlin-Wentzell action functional $S_ T [\psi]$ can be written as
\begin{equation}\label{eq14}
\begin{split}
 S_T [\psi ] &=  \frac{1}{2}\int_0^T {(\dot \psi  - {\bf{F}}_a)\cdot\{a^{-1} [\dot \psi  - {\bf{F}}_a]\} dt} \\
 &=\frac{1}{2}\int_0^T {|\dot{\bf{q}}-{\bf{G}}|^2} dt\\
&\geq\int_0^T{|\dot{\bf{q}}||{\bf{G}}|}dt+W({\bf{B}})-W({\bf{A}})\\
\end{split},
\end{equation}
\noindent where ${\bf{F}}_a=((1-\tau) (u   - u^3  - uv^2), (1+\tau)(- v   - u^2 v))\equiv (F_a^u,F_a^v) $, $\dot{\bf{q}}=(\dot u/(1 - \tau ) , \dot v /(1 + \tau )) $, ${{\bf{G}}(u,v)} = (F_a^u/(1 - \tau ),F_a^v/(1 + \tau ))$ and $\tau\neq 1$. Here, we use $|\dot{\bf{q}}|^2+|{\bf{G}}|^2\geq2|\dot{\bf{q}}||{\bf{G}}|$ to get the inequality. Note that, this inequality is actually an equality, since the DTP $\psi$ minimizes the action functional. The $W(u,v)$ is given by
\begin{equation}\label{eq17}
\begin{split}
W(u,v)&=\int_{}^{(u,v)}{-{\bf{G}}\cdot\dot{\bf{q}}}dt\\
&=\frac{1}{2(1 - \tau) }(\frac{1}{2}u^4  - u^2 ) + \frac{1}{2(1 + \tau)}v^2  + \frac{1}{1 - \tau ^2 }u^2 v^2
\end{split}.
\end{equation}
\noindent The right-hand side of the inequality in Eq.(\ref{eq14}) is a line integral along the directed curve $\psi$, which can be considered as a geometric action functional $\hat S$ similar to the one in Ref.\cite{JCP08061103}. $\hat S=\int_0^T{|\dot{\bf{q}}||{\bf{G}}|}dt =\int_\psi\frac{1}{cos\theta}{\bf{G}}\cdot d\bf{q}$, where $\theta$ is the angle between ${\bf{G}}$ and $\dot{\bf{q}}$. As both ${\bf{G}}$ and $\dot{\bf{q}}$ are always along the $u$ axis, $cos\theta=1$ or $-1$ for any point $(u,v)\in \psi$. Then, we can expand the geometric action functional of the segment from {\bf{A}} to $(u,v)$ near the $u$ axis in powers of $v$ as
\begin{equation}\label{eq18}
\begin{split}
\hat S[\{\psi _0,...,\psi _t =(u,v)\}]
&=\begin{cases}
        |W(u,v)-W({\bf{A}})|  , u\leq0 \\
        |W(u,v)|+ |W({\bf{A}})| ,u>0
   \end{cases}\\
& = m_0(u)+m_2(u)v^2+o(v^2)
\end{split},
\end{equation}
\noindent where $m_2(u)$ is a measure of the transverse stability of the transition path at the point $(u,v)$. If $m_2(u)<0$, any small perturbation in the $v$ or $-v$ direction will lead to a smaller value of the action functional, and the DTP along the $u$ axis will be unstable. Notice that the perturbations in both directions have no difference in decreasing the value of the action functional, which indicates that, there will be two equivalent DTPs, related by $v \to -v$, if $m_2(u)<0$.

For the left segment of $\psi$ where $u\leq0$, we have
\begin{align}
m_2(u)=\begin{cases}
        \frac{1}{2(1 + \tau) }+\frac{1}{(1 - \tau ^2) }u^2,   \tau<1   \\
        -\frac{1}{2(1 + \tau) }-\frac{1}{(1 - \tau  ^2)}u^2,   \tau>1
   \end{cases},
\label{eq19}
\end{align}
\noindent Then, a straightforward calculation shows that the instability region of the on-axis DTP is
\begin{align}
\tau > 1
\label{eq20}.
\end{align}
\noindent Similarly, we find that, for any point with $u>0$, $m_2(u)>0$, which means that small perturbations at the right segment do not affect the stability of the on-axis DTP no matter what value $\tau$ takes.

It is noted that, Eq.(\ref{eq13}) is derived under small delay assumption, and such an expansion has been shown to be valid to quadratic order in $\tau$ \cite{Saa1981}. As Eq.(\ref{eq20}) is not small enough, we can not expect that the derived instability boundary is the exact one. Even so, Eq.(\ref{eq20}) suggests a guiding picture that, there will be a threshold above which the on-axis DTP will undergo a bifurcation via transverse instability on the left segment. What's more, as the geometric action functional $\hat S$ is independent on $T$, the bifurcation of DTP and instability condition Eq.(\ref{eq20}) will also be independent on $T$.

\section{Numerical simulation}

To verify the analytical result, we now derive the DTP by simulation using a recently developed minimum action method \cite{PAM04000637}. The extension of this method to a delayed system is straightforward except for some details. One should note that the delay time defines a upper limit of the step size $\Delta t=T/N$ when we discretize the time domain $[0,T]$ to a $N$-size mesh. To consider the delay effect, $\Delta t$ should be properly chosen so that $\Delta t=\tau /m$, where $m>0$ is a positive integer. We start from a test path $\Phi$ with $\Phi _0 = \bf{A}$ and $\Phi _T = \bf{B}$, and update the path till convergence by iterating on solving the gradient dynamic as
\begin{equation}\label{eq11}
\begin{split}
\frac{{\partial \Phi _t }}{{\partial k}} =& - \frac{{\delta S_T [\Phi ]}}{{\delta \Phi _t }}, 0<t<T\\
\frac{{\partial \Phi _0 }}{{\partial k}} =& \frac{{\partial \Phi _T }}{{\partial k}} = 0
\end{split}.
\end{equation}
\noindent Where, $k>0$ plays the role of pseudo time for the updating, and the action functional of the path $S_T [\Phi]$ can be calculated by the first equality of Eq.(\ref{eq14}) by using $\Phi$ and ${\bf{F}}$ instead of $\psi$ and ${\bf{F}}_a$ respectively. The resulting path(s) with minimum action is(are) our DTP(s) $\psi$, and the rate constant $P$ that the transition from {\bf{A}} to {\bf{B}} occurs can be approximately estimated by
\begin{align}
P \asymp \mathop {\lim }\limits_{T \to \infty } \exp \{  - \frac{1}{\varepsilon }S_T [\psi ]\},
\label{eq6}
\end{align}
\noindent where $f(\varepsilon ) \asymp g(\varepsilon )$ if $\log f(\varepsilon )/\log g(\varepsilon ) \to 1$ as $\varepsilon \to 0$. When the DTP bifurcates, $P$ is calculated via summing ones of each path.

\begin{figure}
\centerline{\includegraphics*[width=0.7\columnwidth]{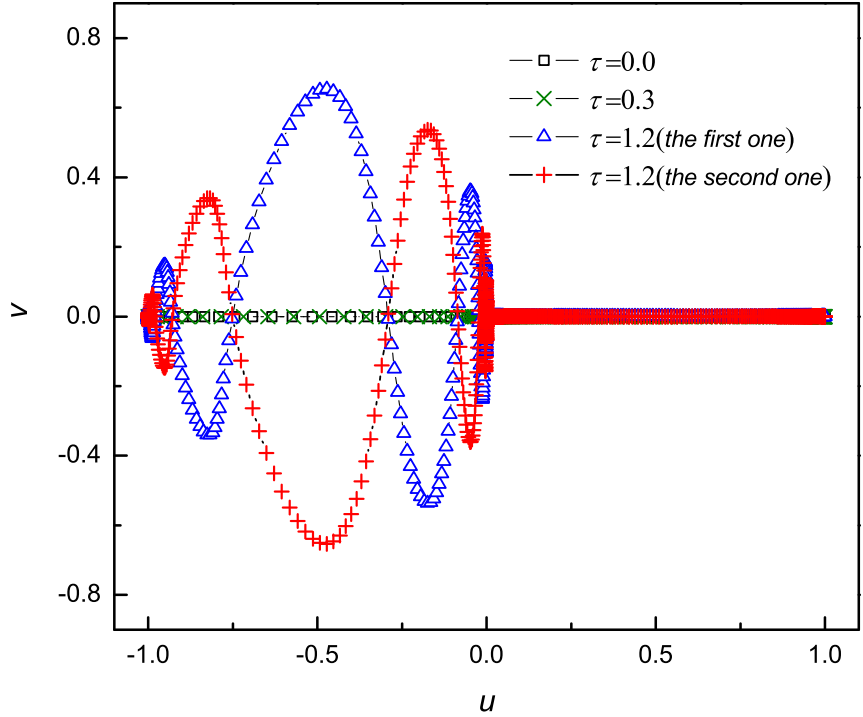}} \caption{The dominant transition pathways of delayed Maier-Stein model when $\tau =0.0, 0.3$ and $1.2$. \label{Fig4}}
\end{figure}

\begin{figure}
\centerline{\includegraphics*[width=0.7\columnwidth]{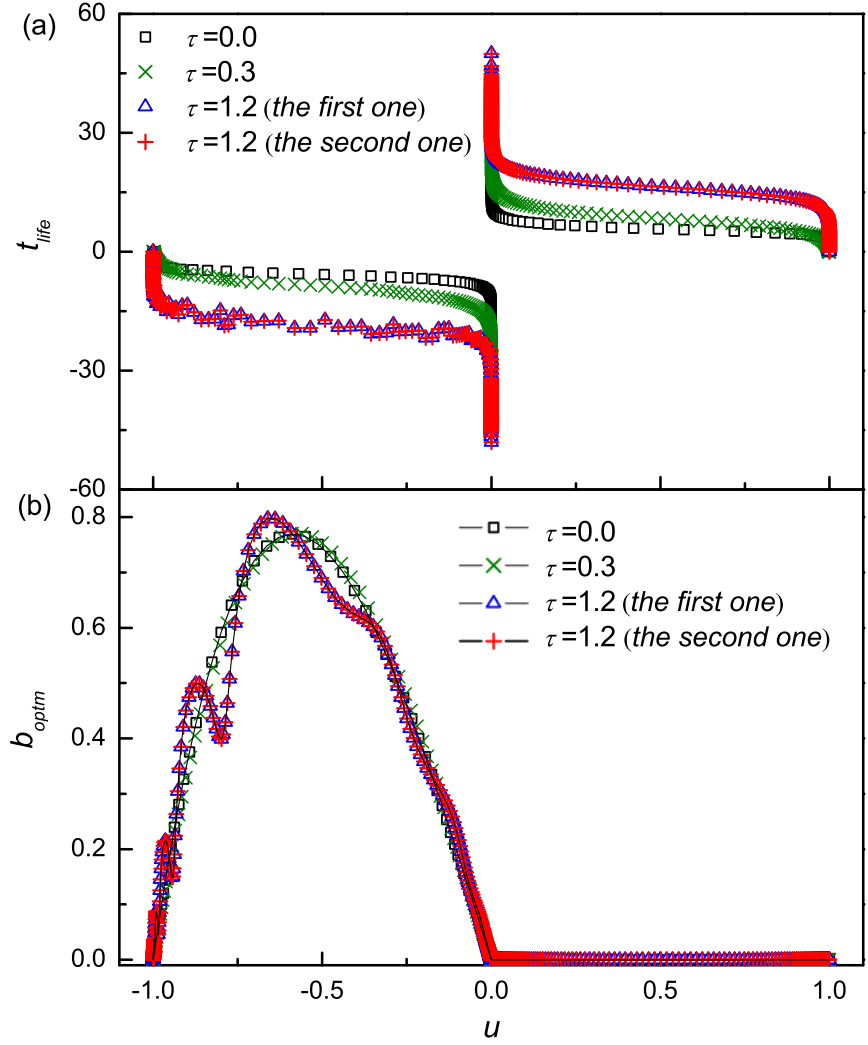}} \caption{(a) The life time $t_{life}$ and (b) the optimal fluctuational force $b_{optm}$
for each point of the dominant transition pathways shown in Fig.\ref{Fig4}. \label{Fig5}}
\end{figure}

The numerical DTPs for $\tau =0.0, 0.3$ and $1.2$, and $T=100$ are plotted in Fig.\ref{Fig4}. It can be observed that two equivalent DTPs, symmetric under $v \to -v$, appear, which means that, the DTP does undergo a bifurcation in the trajectory space when $\tau$ is large enough. When $\tau =0.3$, only one on-axis DTP exists. While $\tau =1.2$, two off-axis DTPs are observed. For further quantitative analysis of the DTPs, we define the life time, as defined by X. Zhou, etc.\cite{CMS10000341}, to be the (signed) time in the deterministic system starting from a given position to reach the $\kappa$-neighborhood of {\bf{A}} (negative life time) or {\bf{B}} (positive life time), where $\kappa  = 10^{ - 5} $ in our simulation. A negative life time indicates that the point is in the attraction basin of {\bf{A}}, and a positive one implies that the point will be attracted to {\bf{B}}. Thus, the transition state $\psi_{tran}$ can be determined by the middle of the last point with negative life time and the first point with positive life time. The life times $t_{life}$  for each point on the numerical DTPs are presented in Fig.\ref{Fig5}(a). We find that, all the DTPs are separated by the origin (0,0) into two segments which lie in basins of attraction of {\bf{A}} and {\bf{B}}, respectively. The fluctuation of the negative life time at $\tau = 1.2$ may due to the fact that, the trajectory back to {\bf{A}} doesn't overlap with the DTP in off-axis region, and it is easier for some point to return to {\bf{A}} in the deterministic delayed system. In addition, we have also calculated the amplitude of optimal fluctuation forces $b_{optm}=\left| {\dot \psi_t  - {\bf{F}}(\psi_t ,\psi_{t-\tau}  )} \right|$ for each point on the numerical DTPs. The forces are considered to be optimal is due to the fact that, as they are calculated along the dominant transition pathway $\psi$, we will get this dominant transition pathway back by applying these forces to Eq.(1). Since $b_{optm}$ is proportional to the deterministic force {\bf{F}} which is 0 at {\bf{A}} and the transition state, its value will be near 0 for the points close to {\bf{A}}, then increases and reaches its maximum at some middle point, and decreases to nearly 0 for the points close to the origin, as presented in Fig.\ref{Fig5}(b). Besides of this, several points can be addressed. Firstly, for all $\tau$, strong fluctuation force is needed for the system to escape from the attraction of {\bf{A}}. Just after the system passes through the origin,  $b_{optm}$ decreases to zero, immediately. Secondly, the curve at $\tau=0.3$ is overlapped with its analog at $\tau=0.0$, which confirms that the on-axis DTP is still stable when the delay time is small. Finally, for $\tau=1.2$, $b_{optm}$ of the left segments show large discrepancies from the one at $\tau=0.0$ obviously, but the right segments do not. The results given by Fig.\ref{Fig4} and Fig.\ref{Fig5} are consistent with our analytical results.

\begin{figure}
\centerline{\includegraphics*[width=0.5\columnwidth]{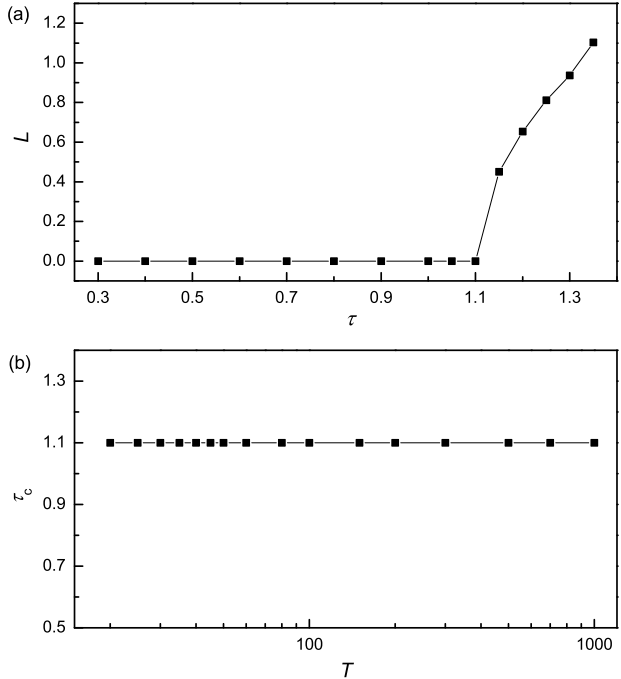}} \caption{The bifurcation of the dominant transition pathway. (a) The maximal distance $L$ from dominant transition pathway to the $u$ axis as a function of delay time $\tau$. $L$ arises from nearly zero to a remarkable value while delay time pass through the threshold $\tau _c \approx 1.1$. (b) The scaling of threshold $\tau _c$ with observation time $T$. \label{Fig6}}
\end{figure}

In order to quantitatively describe the bifurcation of DTP, we introduce the maximal distance from DTP to $u$ axis, $L$, as follows
\begin{align}
L = \max (l_t ), 0\leq t \leq T
\label{eq21}
\end{align}
\noindent where $l_t$ is the distance from $\psi _t$ to the $u$ axis. In Fig.\ref{Fig6}(a), $L$ as a function of delay time $\tau$ is shown. It can be seen that $L$ stays nearly zero for small delay time until $\tau\geq\tau _c\approx 1.1$. We note that the threshold $\tau _c$ is close to the result given by Eq.(\ref{eq20}). To make sure the bifurcation is not a result of small $T$ in our simulation, we test the scaling behavior of $\tau _c$ with $T$, which is shown in Fig.\ref{Fig6}(b). The independence of $\tau _c$ on $T$ implies that the bifurcation of DTP in trajectory space does occur for  large $T$s, which also confirms our analysis. For infinite $T$, it is not available to calculate DTP directly (A geometric minimum action method has been developed by Heymann et. al.\cite{PRL08140601} to deal with the infinite $T$ problem, however, it is not suitable for delayed systems).

\section{Discussion}
\begin{figure}
\centerline{\includegraphics*[width=0.5\columnwidth]{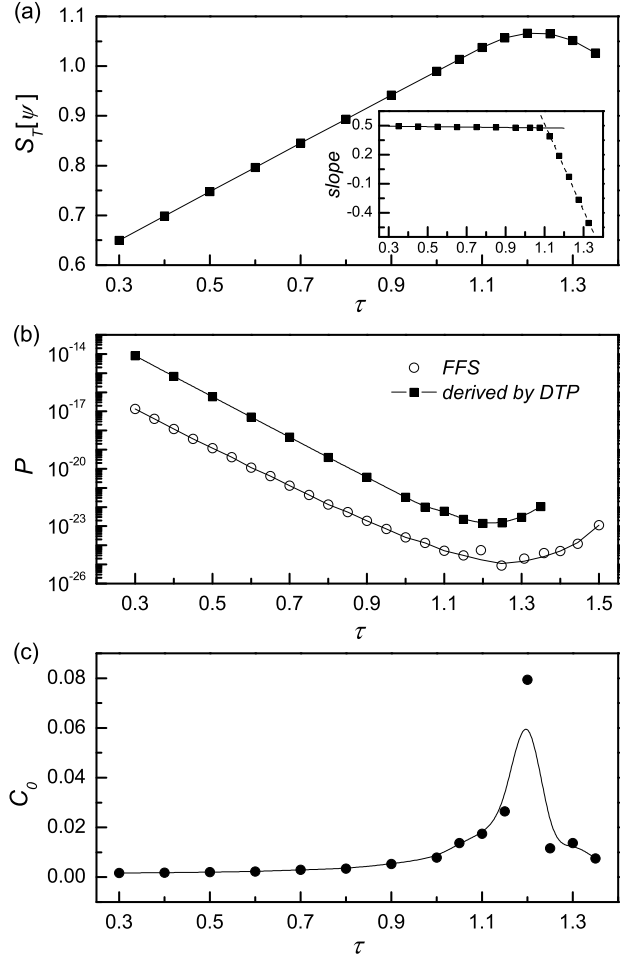}} \caption{(a) The minimal action functional $S_T [\psi ]$, (b) transition rate constant $P$ and (c) its prefactor $C_0$ as a function of delay time $\tau$. The insert panel of (a) is the slope $\delta S _T [\psi]/\delta \tau$ as a function of delay time $\tau$. Here, $\varepsilon=0.02$ for the forward flux sampling (FFS) simulation.\label{Fig8}}
\end{figure}

An important quantity derived from the DTP is the action functional $S _T [\psi]$ which can be calculated by the first equality of Eq.(\ref{eq14}) by apply $\bf{F}$ instead of ${\bf{F}}_a$ and is related to the transition rate constant by Eq.(\ref{eq6}). $S _T [\psi]$ as a function of $\tau$ is shown in Fig.\ref{Fig8}(a). When $\tau$ is small, $S _T [\psi]$ increases almost linearly as $\tau$ increases, however, when $\tau$ is large,  $S _T [\psi]$ departs from the linear relationship obviously. To show this clearly, we plot the slope $\delta S _T [\psi]/\delta \tau$ in the insert of Fig.\ref{Fig8}(a). For $\tau<1.1$, $(\delta S _T [\psi]/\delta \tau)/\tau \approx 0.0$ which means the slope is nearly unchanged. When $\tau>1.1$, the slope decreases as $\tau$ increases at a rate of $(\delta S _T [\psi]/\delta \tau)/\tau \approx -4.5$. For comparison, we also run the dynamic equation Eq.(\ref{eq12}) directly involved with forward flux sampling approach (FFS)\cite{PRL05018104} to get the transition rate constant $P$ at different $\tau$. $P$ at $\varepsilon=0.02$ are plotted in Fig.\ref{Fig8}(b). The nontrivial dependence of $P$ on $\tau$ is also observed, which is consistent with the result derived by DPT qualitatively. The quantitative difference is due to the fact that, while we estimate $P$ by Eq.(\ref{eq6}), the crossing through the transition state is considered as ballistic, i.e., it assumes that every crossing gives rise to a successful transition. For the diffusive crossing, some crossings may turn back to {\bf{A}} and $P$ is overestimated. The true $P$ should include a prefactor $C_0$, which evaluate the ratio of successful transition to total crossing,  as well as the exponential factor Eq.(\ref{eq6}). Since FFS simulation samples the transition rate constant directly by system's dynamics, we can calculate $C_0$ by the ratio between the two curves in Fig.\ref{Fig8}(b). Fig.\ref{Fig8}(c) presents $C_0$ as a function of $\tau$. The abrupt increasing of $C_0$ near the bifurcation point $\tau_c$ implies that the system may undergo some sort of critical behavior as a result of delay induce DTP bifurcation.

%\begin{figure}
%\centerline{\includegraphics*[width=1.0\columnwidth]{Figure5.eps}} \caption{(a) The minimal action functional $S_T [\psi ]$ as a function of delay time $\tau$. The insert
%panel: the slope as a function of delay time $\tau$. (b) The transition rate constant $P$ as a function of delay time $\tau$. Here, $\varepsilon=0.02$.\label{Fig8}}
%\end{figure}

\begin{figure}
\centerline{\includegraphics*[width=0.5\columnwidth]{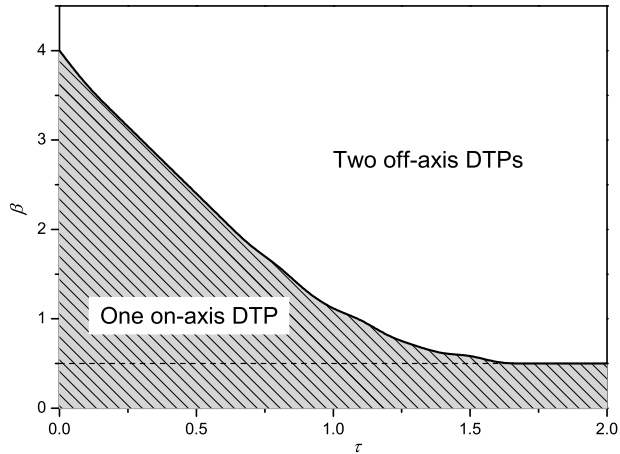}} \caption{The bifurcation diagram of dominant transition pathways in $\tau-\beta$ plane.\label{FigBD}}
\end{figure}

In the above simulation, we have fixed $\beta=1$ to avoid the influence of non-conservative effects other than time delay. It has been reported that, bifurcation of the DTP also occurs when $\beta$ is varied \cite{PRL93001783}. To understand the dependence of the DTP on the both parameters $\tau$ and $\beta$, a two-dimensional bifurcation diagram in the $\tau-\beta$ plane can be plotted. By extensive MAM simulations at different $\beta$ and $\tau$, we plot this diagram in Fig.(\ref{FigBD}). It seems that, the bifurcation curve asymptotically approaches the line $\beta=0.5$, i.e., the symmetry breaking bifurcation of DTP doesn't occur for any delay time when $\beta<0.5$. We have tried to figure out the underlying mechanism theoretically, however, a potential-like quantity similar to Eq.(\ref{eq17}) is not available for the non-conservation case $\beta\neq1$, which may need further research.
%\section{conclusion\label{section5}}
\section{Conclusion}
In summary, the dominant transition pathway between two metastable states of a delayed stochastic system is studied. To show the effect of delay, we consider a modified version of the Maier-Stein model with linear delayed feedback as an example, whose original model is a typical symmetric {\lq double well\rq} system, to study the noise induced transition. Our analysis by small delay approximation shows that, time delay will induce a new DTP via a bifurcation in trajectory space when the delay time passes through a threshold $\tau_c$. By employing a recently developed minimum action method, we can calculate the DTP by minimizing the Freidlin-Wentzell action functional for transition path, numerically. The numerically derived DTP confirms that the bifurcation does occur for $\tau >\tau_c=1.1$. Other details of DTP bifurcation are also verified by numerical results, including bifurcation via transverse instability, bifurcation at left segment, arising of two equivalent DTPs after bifurcation, etc. From the DTP, the transition rate constant can be derived, which shows distinct dependence on the delay time below and above the threshold. This dependence is also observed by directly running the dynamic. The bifurcation diagram is also investigated. Since time delay is an important factor in many real systems, we believe that the present study can shed new light on understanding the mechanism of fluctuation-driven transitions in experimental studies and open more perspectives on the study of fluctuation-driven phenomena in non-conservation systems.

\acknowledgments
This work is supported by National Science Foundation of China (21125313, 20933006, 91027012).

%\bibliography{DTP}
%\bibliographystyle{apsrev}

\end{CJK*}
\end{document}